# Modified Kedem–Katchalsky equations for osmosis through nano-pore

Liangsuo Shu[1,3], Xiaokang Liu[1], Yingjie Li[2], Baoxue Yang[2], Suyi Huang[1], Yixin Lin[1,4], Shiping Jin[1,3,4,*]

[1]School of energy and power engineering, Huazhong University of Science & Technology. Wuhan, China.
[2] State Key Laboratory of Natural and Biomimetic Drugs, Department of Pharmacology, School of Basic Medical Sciences, Peking University, Beijing, China;
[3]Innovation Institute, [4]China-Europe Institute for Clean and Renewable Energy, Huazhong University of Science & Technology. Wuhan, China.

Address correspondence to:

**Shiping Jin**

Innovation Institute, School of Energy and Power Engineering,

Huazhong University of Science & Technology.

1037 Luoyu Road, Wuhan, 430074, China;

E-mail: jinshiping@hust.edu.cn

Modified Kedem–Katchalsky equations for osmosis through nano-pore

## Abstract

This work presents a modified Kedem–Katchalsky equations for osmosis through nano-pore.

$$J_v = L_p(\Delta P - \sigma_o \Delta\pi)$$

$$J_s = \omega L_p \Delta\pi + x(1-\sigma_s)cJ_v$$

$\sigma_o$ of a solute was found to be chiefly affected by the entrance of the pore while $\sigma_s$ can be affected by both the entrance and the internal structure of the pore. Using an analytical method, we get the quantitative relationship between $\sigma_o$ and the molecule size. The model is verified by comparing the theoretical results with the reported experimental data of aquaporin osmosis. Our work is expected to pave the way for a better understanding of osmosis in bio-system and to give us new ideas in designing new membranes with better performance.

*Keywords*: K–K equations, osmotic coefficient, aquaporin

## 1. Introduction

The osmotic water fluxes coupled to solute transport is common in both biology systems and membrane technology: ordered mass transfer across the cell membrane plays important physiological roles; different kinds of artificial membranes have been applied to many areas such as desalination[1–3], purification of biotechnology products[4–7] and capturing osmotic energy[8–13]. Kedem-Katchalsky (K-K) equations[14] and their improved versions[15] are usually used in analyzing such phenomena. The effects of solute molecular size and shape on the reflection coefficients in K-K equations have been an important research topic because of its definite application background in both bio-osmosis and membranes technology. In this topic, the long pores ($L>>D$, the length of the pore is much greater than its diameter) had been systemic studied[16–18]. For long pore, the entrance effects are usually neglected as an approximation. However, many researches have shown that[19–22] the entrance effects play very important role in mass transfer through short channels. What's more, a study by Sisan and Lichter revealed that the entry effects should not be neglected even for long pores at the nanoscale[23].

Although widely used, it is difficult to make a direct verification of this model by experiments because the pores of most artificial membranes are inhomogeneous in both sizes and shapes. The discovery of aquaporin (AQP) provides us a perfect touch-stone to check the K-K equations. All channels of one kind of AQP are the expressions of the same gene and thus have exactly the same structure, which can be determined in high resolution by both electron crystallography of 2D crystals and X-ray crystallography of 3D crystals [24,25]. Using stopped-flow[26], the fast non-equilibrium osmosis process through AQPs can be recorded. AQP1 is the main water channel of red blood cell (RBC) and contributes most of its water permeability. AQP1 has a selectivity filter which is so narrow that only water molecule can pass through[24]. Therefore, from the K-K equations, the $\sigma$ of a solute to RBC should be a value closed to 1. However, many experiments about RBC indicate $\sigma$s of some small molecules are much smaller than 1 and have a close relation with their molecular sizes[26–28]. This phenomenon had been qualitatively explained by Hill and coworkers as a result of that these impermeable solutes can enter the vestibules of AQP1 to contribute parts of their momentums[28]. In this way, the two $\sigma$s of K-K equations can be different with each other. They also defined two coefficients, the "osmotic" coefficient ($\sigma_o$) and the "ultrafiltration" coefficient ($\sigma_s$)[29] to replace the reflection coefficient($\sigma$) in K-K equations. In a study of the effects of molecular shape on reflection coefficient by Bhalla and Deen[18], they proposed a similar idea (osmotic reflection coefficient and filtration reflection coefficient).

Based on the summary of the existed researches, we provide a modified model of K-K equations for nano-osmosis.

$$J_v = L_p(\Delta P - \sigma_o \Delta \pi) \quad (1)$$

$$J_s = \omega L_p \Delta \pi + x(1 - \sigma_s) c J_v \quad (2)$$

where $J_v$ is the volume flux, $J_s$ is the solute flux, $\Delta P$ and $\Delta \pi$ are the pressure and osmotic pressure differences on both sides of the semipermeable membrane, respectively, $L_p$, is the hydraulic conductivity, $\omega$ is the solute permeability, $x$ is the selectivity rate of the selectivity filter of the pore, $c$ is the solute concentration outside the pore.

Although the qualitative analysis by Hill reveals the nature of the problem about the defects of K-K equations when applying to nano-osmosis, the qualitative explanation also limits its application. In this work, a quantitative result and its possible applications are presented. By analyzing the experimental data[26–28,30], $\sigma_o$ of a solute was found to be chiefly affected by the entrance of a pore while $\sigma_s$ can be affected by both the entrance and the internal structure of the pore. Using an analytical method, we get the quantitative relationship between $\sigma_o$ and the molecular size as follows:

$$\sigma_o = \begin{cases} \beta^3 & 0 < d_m \le d \\ 1 & d_m > d \end{cases} \tag{3}$$

where $d$ is the effective diameter of entrance of the one pore, $d_m$ is the effective diameter of a solute molecule, and $\beta = d_m / d$, is the relative size of a solute molecule to the pore. Equation (2) describes how the molecular sizes of solutes affect the osmotic pressure: the $\sigma_o$s of molecules increase with a third power of their $\beta$s up to a constant value 1(Fig. 2). This phenomenon is suggested to be named as *"Solomon-Hill effect"* because that it was first measured by Solomon[26,27,31] and first explained by Hill[28]. Solomon found this phenomenon in studying water permeability of red cell and predicted the existence of water channel (now known as Aquaporin). Equation (3) gives a quantitative description of *Solomon-Hill effect.* From the derivation process below, we will find that *Solomon-Hill effect* is a result of the interaction between solute molecule and the entrance of a pore. Therefore, Equation (3) remains valid even for solutes which cannot pass through a pore.

## 2. Method and result

### 2.1. "gas analogy" of Van't Hoff, McMillan and Mayer statistical theory for solution

The analytical method used in the paper might be associated with early "gas analogy", which was first introduced by Van't Hoff[32] to give an explanation for his equation of osmotic pressure. However, his gas analogy had been questioned as it seems to ignore the intermolecular forces in liquid and once abandoned for a long time[33]. After six decades, McMillan and Mayer[34] proposed the statistical theory of the osmotic pressure and gave the theoretical explanation basis for the classic gas analogy of Van't Hoff: the presence of the solvent only implicitly appears as the potentials of average force of the solute molecules, which are influenced by the presence of the solvent[34]. For a dilute solution, compared with their mean free path, the mean distance between the solute molecules is so large that the interaction between them is weak. In this situation, the solvent can be regarded as a continuous medium, acting as the space of "idea gas" of solute molecules. Thus, the result of McMillan and Mayer statistical theory for osmotic pressure of dilute solution can be simplified to the Van't Hoff Equation.

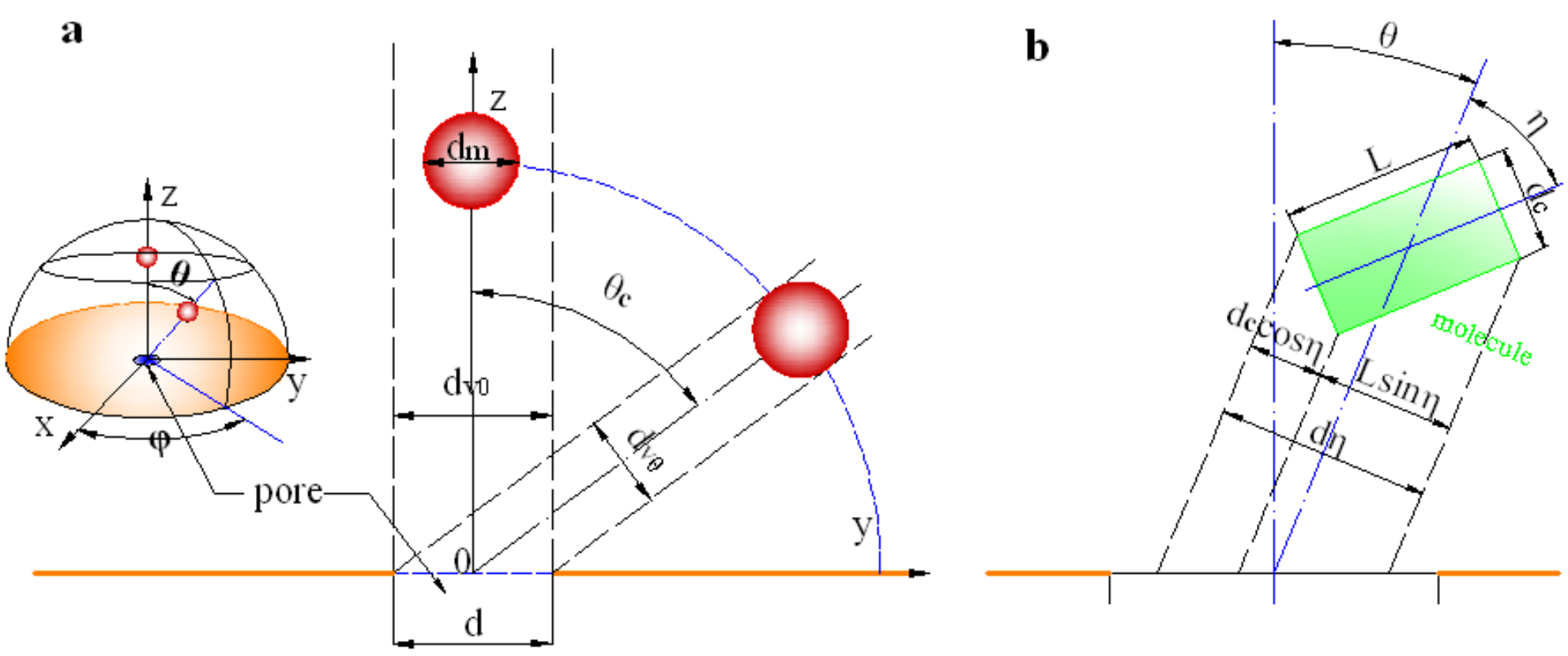

Fig. 1. Diagram of our model. a, the supercritical collision of a solute molecule, a small three-dimensional model and the two-dimension magnified view. b, the equivalent diameter of a small molecule.

### 2.2. $\sigma_o$ and Solomon-Hill effect

Fig. 1a is a schematic of the model. The visible diameter of the pore ($d_{V\theta}$) varies with the direction ($\theta$). When $\theta$ is 0, the visible diameter ($d_{V0}$) becomes a maximum and is equal to the diameter of the pore $d$. Considering a molecule with an effective diameter ($d_m$), the molecule can just enter the pore ($d_m = d_{V\theta}$) along with a critical direction $\theta_c$, the cut-off angle of the molecule to the pore.

$$\theta_c = \arccos \beta \quad (4)$$

where $\beta = d_m / d$, is the relative size of a molecule to the pore. If a molecule moves along a direction larger than $\theta_c$, it will be larger than the visible diameter of the pore ($d_m > d_{V\theta}$) and won't be able to enter the pore, and consequently will not contribute its momentum. This kind of collision is called a supercritical collision. Conversely, all the collisions of solvent molecules can contribute their momentum to pore. Therefore, compared with a pure solvent, there will be a negative pressure for the pore resulting from the supercritical collisions of solute molecules. This negative pressure is the osmotic pressure.

From the Knudsen cosine law[35], the number of molecules ($dN$) that move to a finite area ($dA$) along the direction $\theta$ in unit time ($d\tau$) can be described as follows:

$$dN = \frac{n\overline{\omega}}{4\pi} \cos\theta d\Omega dA d\tau \quad (5)$$

where $N$ is the number of solute molecules and $V$ is the volume, $n = N/V$ is the number density per unit volume of this kind of molecule, $\overline{\omega}$ is the root-mean-square velocity of gas molecules, $d\Omega = \sin\theta d\theta d\varphi$ is the solid angle between $\theta$ and $\theta + d\theta$.

The vertical component of the momentum change ($\Delta m\overline{\omega}$) of a molecule from the direction $\theta$ is $2m\overline{\omega}\cos\theta$, where $m$ is the weight of the molecule. The negative pressure of the pore can be calculated as follows:

$$\pi dAd\tau = \Delta m\overline{\omega} \times dN = 2m\overline{\omega}\cos\theta \times \frac{n\overline{\omega}}{4\pi}\cos\theta d\Omega dAd\tau \qquad (6)$$

After simplifying and integrating the above equation to get the sum of supercritical collisions,

$$\pi = \int_{\Omega_c}^{2\pi} \frac{2nm\overline{\omega}^2}{4\pi}\cos^2\theta d\Omega = \frac{nm\overline{\omega}^2}{3}\cos^3\theta_c \qquad (7)$$

From the equipartition theorem, we know that

$$\overline{\varepsilon} = \frac{1}{2}m\overline{\omega^2} = \frac{3RT}{2N_A} \qquad (8)$$

where $\overline{\varepsilon}$ is average molecular kinetic energy, $N_A$ is Avogadro's number. Employing equations (4), (7) and (8), we get

$$\pi = \sigma_o cRT \qquad (9)$$

where $\sigma_o = \beta^3$ is the osmotic coefficient, $c = n / N_A$ is the molar concentration of the solute molecules in solution. If the solute molecule is larger than the pore ($d_m > d$), the molecule contributes no pressure to the pore, so we can get equation (3) at last.

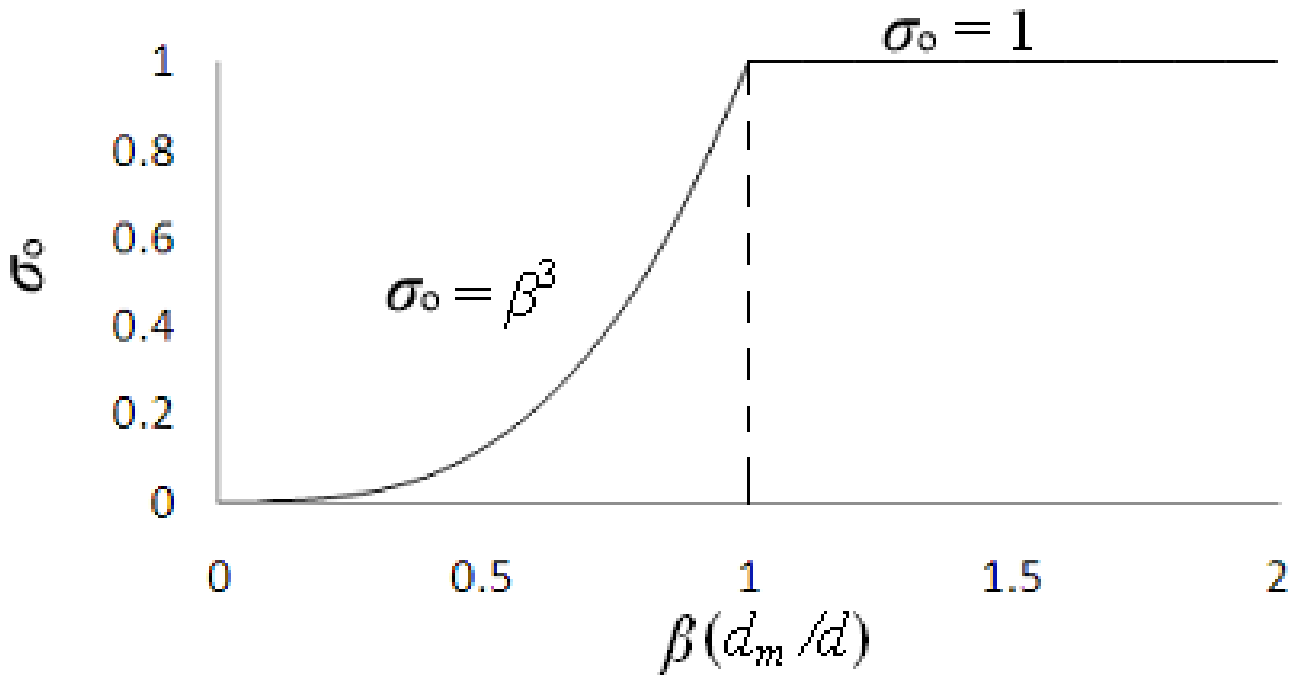

Fig. 2. A quantitative description of *Solomon-Hill effect,* the effect of solute molecular size on the osmotic coefficients ($\sigma_o$). $\sigma_o$ is a function of solute molecular size to a pore $\beta(d_m/d)$: $\sigma_o$ increases with a third power of $\beta$ up to a constant value 1.

### 2.3. $\sigma_s$ and primary screening effect

If the size of a solute molecule is smaller than the entrance of a pore ($d_m<d$), the probability that it can pass through the pore is affected by both the entrance and the internal structure of the pore. The entrance offers a primary screening effect: the solute molecule which moves along a direction larger than the critical cut-off angle will be kept away. By integrating Equation (5) from $\theta_c$ to 2π, we can get

$$\sigma_s = \beta^2 = \sigma_o^{2/3} \tag{10}$$

There are two "fates" for the solute molecules entering the pore: passing or being repatriated, which is determined by the selectivity filter in the pore. The passing rate ($\alpha$) can be described as the follow equation

$$\alpha = x(1-\sigma_s) \tag{11}$$

where $x$ is the selectivity rate of the selectivity filter in the pore.

The number of a solute that can enter the pore ($N$) in unit time is

$$N = \alpha n \pi d^2 \frac{\sqrt{3RT}}{16\sqrt{M}} \tag{12}$$

where $M$ is the molar mass of solute, and $n$ is its number density. Equation (12) indicates that lighter and smaller molecules have a better chance of passing through a pore.

## 3. Comparing with experimental data

In order to verify the accuracy of the model, we compared its theoretical predictions with experimental results. The theoretical predictions agreed well with experimental results, showing high accuracy of our model.

### 3.1.Effective diameter of small molecule

In the basic model above, the solute molecule was regarded as spherical to obtain the fundamental relation between $\sigma_o$ and $d_m$. However, besides special cases, most solute molecules are non-spherical. So, how to define their equivalent diameters?

Most small molecules can be regarded as short columns, the geometrical shape of which can be described by length ($L$) and cylindrical diameter ($d_c$). in the qualitative discussion by Hill[28], cylindrical diameters were chosen as the indicator of molecular size. However, the thermal motion of a molecule in solution is so intense that one molecule may impact a pore in any possible posture. As shown in Fig.1 (b), a molecule moves along an arbitrary direction $\theta$ with an attitude angle ($\eta$, the angle between its axis of symmetry and its flying direction). The viewed diameter of the molecule from the pore ($d_\eta$) can be described by

$$d_\eta = d_c \cos\eta + L\sin\eta \tag{13}$$

Integrating the above equation at the value range of $\eta$ from 0 to π/2, the equivalent diameter of a small molecule can be calculated,

$$d_m = \frac{1}{\pi/2}\int_0^{\pi/2}(d_\eta)d\eta = \frac{2(d_c + L)}{\pi} \tag{14}$$

For big molecules, their Stokes radius can be used as their effective diameter.

### 3.2. $\sigma_o$ of small molecule to AQP1

The water permeability of a cell is usually a joint contribution from its cytomembrane of lipid bilayer and one or more kinds of water channels. The apparent osmotic coefficient $\sigma^*$ of a cell can be described as:

$$\sigma^* = \sum_{i=0}^{n} k_i \sigma_i \tag{15}$$

where $k_i$ is the proportion of $i$th water channel with $\sigma_i$, while $i$ = 0 stands for the lipid bilayer. Assuming that AQP and the lipid bilayer contribute the total water permeability of a cell, there will be only two terms at the right of equation (15):

$$\sigma^* = k_0 + k_1\sigma_1 = \begin{cases} k_0 + k_1 \dfrac{d_m^3}{d_1^3} & 0 < d_m \le d \quad (1) \\ k_0 + k_1 = 1 & d_m > d \quad (2) \end{cases} \tag{16}$$

where $k_0$ and $k_1$ are, respectively, the contributions of the lipid bilayer and of AQP to the total osmotic water permeability.

For human RBC, AQP1was found to contribute more than 85% of the total osmotic water permeability and only about 10% of lipid bilayer permeation[36] at 20°C. Toon & Solomon[27] measured the $\sigma_o$s of small solutes to human RBC at 25°C and their data was shown in Fig. 3a. All molecular lengths and cylindrical diameters were taken from ref. [37]. The calculated $k_0$ =11.3% and $k_1$ =88.7%, conforming well to the results of ref. [36]: $k_0$ ≈10% and $k_1$ >85%. Some experiments[38,39] with UT-B (Urea Transporter-B) indicated that UT-B also contributes a part of the water permeability of the erythrocyte. The experimental methods using RBC[27,36] were unable to distinguish AQP1 from UT-B. Their results regarding AQP1 were actually an integrative effect of AQP1 and UT-B. In our previous work studying the information thermodynamics of AQP1 osmosis, the RBCs of both common mice and UT-B knockout mice were used[40]. The influence of UT-B to $\sigma_o$ of small molecule to RBC was found to be limited.

Fig. 3b shows the measured $\sigma_o$ of HA/FA-AQP1[30] expressed in Xenopus oocytes. Same molecular sizes were obtained using MMsINC[41], a database that gives molecules' Sterimol sizes. The lengths were taken from the Sterimol $L$ directly, while cylindrical diameters were the sum of Sterimol B1 and Sterimol B4. $\sigma$s of all small solutes to AQP1 expressed in Xenopus oocytes were about 1. The intercept for HA/FA-AQP1 indicated the contribution of water permeability by the cell membrane of oocytes was 5.7%, agreeing well with the measured result 5.2% in ref. [30]. The variance between $\sigma_o$ to erythrocytes AQP1[27,28,42] and $\sigma_o$ to oocytes-expressing AQP1[30,43] shows that there is a difference in vestibule structure of AQP1 when it expresses at different kinds of membranes. The studies on the lipid–protein interactions of AQP1 [44–47] suggest that it is a result of the effects of membrane tension.

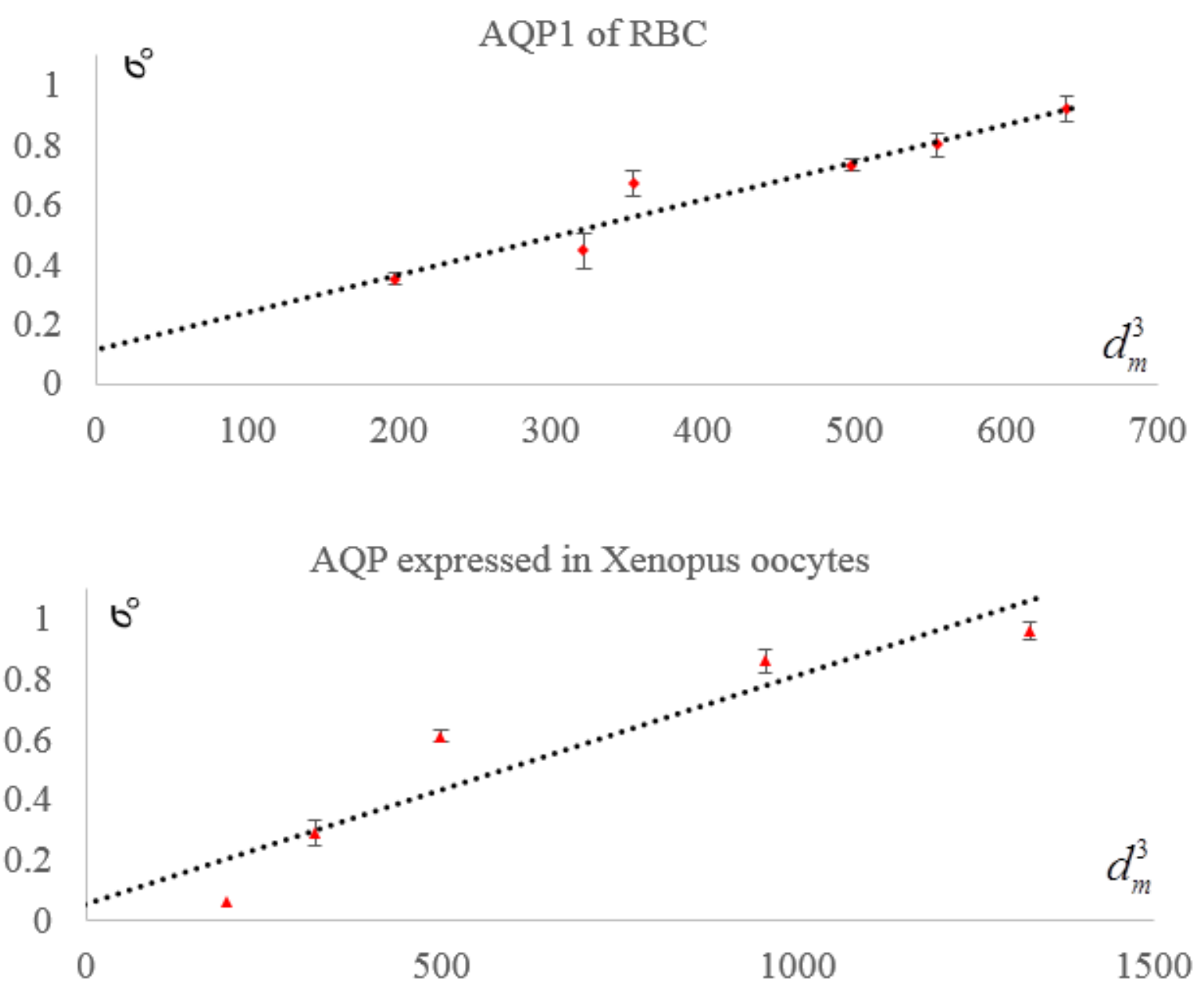

Fig.3. Experimental data for the osmotic coefficients ($\sigma_0$) of small solutes to red blood cell AQP1 (a,) and to AQP expressed in Xenopus oocytes (b,). Two straight lines are linearization of the experimental data ($R^2$=0.92 for a and $R^2$=0.88 for b). The intercepts are the percentage of the water permeability by cell membranes. In a, this value is 11.3%, which agrees well with the experimental result of ref. [17], $\approx$10%; In b, this value is 5.7%, which also agrees well with the experimental result in ref. [12], 5.2%. HA/FA - AQP1 is a double mutants of AQP1 (H180A/F56A).

### 3.3.Primary screening effect of the vestibules of AQPs

The vestibules of AQPs can offer a primary screening to different molecules before their constriction region. This primary screening effect, to a certain extent, can protect their excellent osmotic water permeability by preventing possible channel blockage of AQPs. Fig. 3 shows a comparison between the theoretical values calculated from $\sigma_o$ using equation (10)-(12) (assuming all the selectivity rates ($x$s) of permeable solutes is a constant approximatively) and the experimental values[30] with an index of the ratio of solutes permeability to the water permeability ($P_s/P_f$). Water permeability ($P_f$) is a good indicator of the largest potential of the permeability of a channel. Therefore, $P_s/P_f$ can characterize the degree of difficulty for a solute to pass through a pore. Both of these results were presented by using a non-dimensional method for easier comparison. The error in theoretical value was a propagation of the experimental errors of $\sigma_o$s. The slope of the linearized line is 0.96 with a correlation coefficient of 0.95.

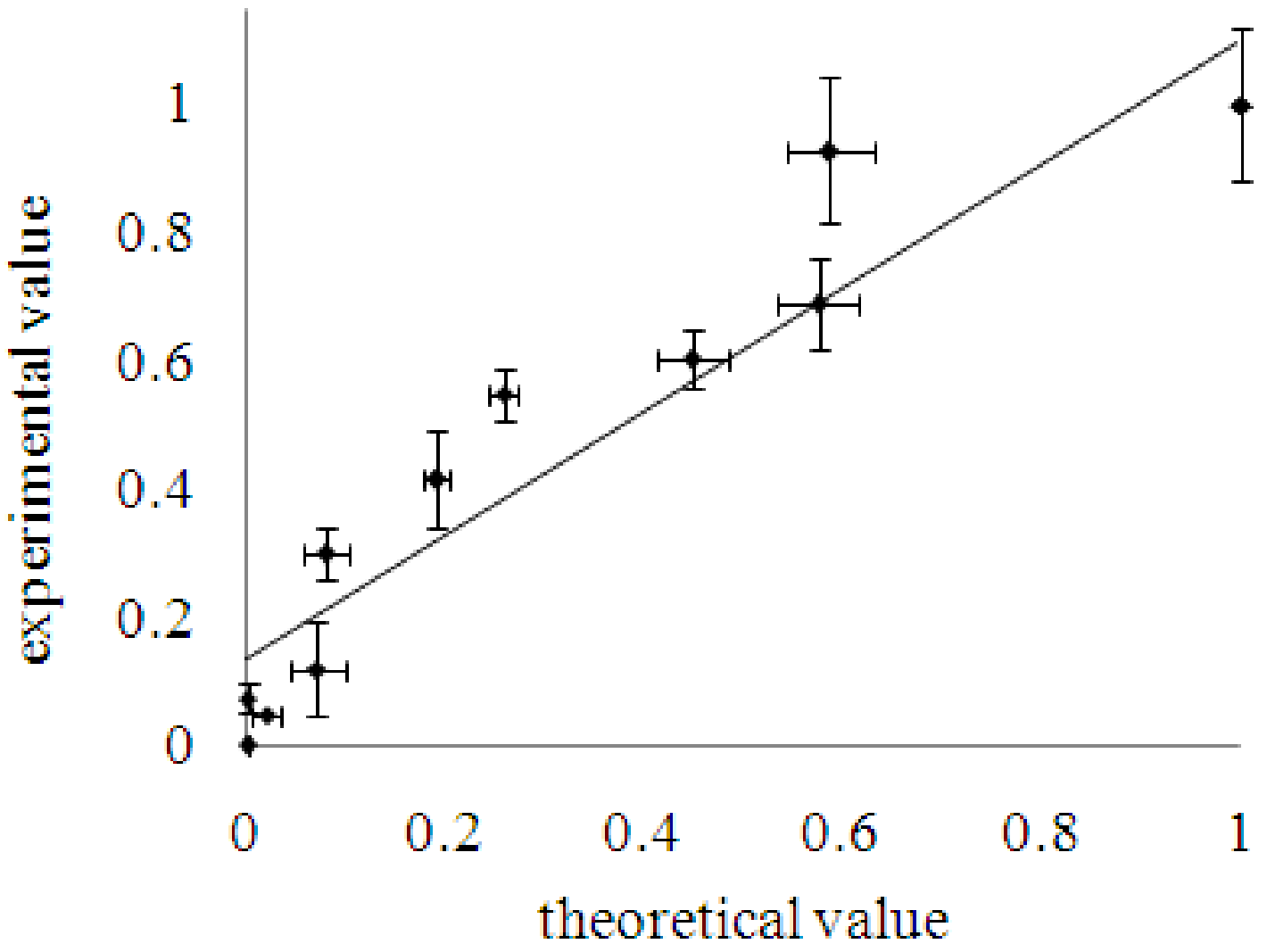

Fig. 4. The primary screening effect of AQP, a comparison between the theoretical value calculated by equation (7) and the experimental value of reference[30], involving

HA/FA-AQP1, SH-AQP9 and SH/GI-AQP9. The slope of the linearized line is 0.96 with a correlation coefficient of 0.95. In the calculation of the theoretical value, all the selectivity rates ($x$s) of permeable solutes is approximatively regarded as a constant, which is found to be reasonable from the results. However, we can still find that some solutes are easier to pass through AQP compared with other ones (their experimental values are greater than theoretical values and should have bigger selectivity rates).

## 4. Conclusion and discussion

In this work, we provided the quantitative relationship between $\sigma_o$ and the molecule size, which is verified by comparing with the experimental data of aquaporin osmosis. The K-K equations are improved to extend its application from marco-osmosis to nano-osmosis.

$$\begin{cases} J_v = L_p(\Delta P - \sigma_o \Delta\pi) \\ J_s = \omega L_p \Delta\pi + x(1-\sigma_s)cJ_v \end{cases}$$

$\sigma_o$ represents the momentum loss because of the shelter of the entrance of the pore, while $\sigma_s$ is the proportion of molecules which are kept out because of primary screening effect. Therefore, we think *primary sieving coefficient* is a more appropriate name for $\sigma_s$.

$$\sigma_o = \begin{cases} \beta^3 & 0 < d_m \le d \\ 1 & d_m > d \end{cases}$$

$$\sigma_s = \begin{cases} \beta^2 & 0 < d_m \le d \\ 1 & d_m > d \end{cases}$$

The passing rate ($\alpha$) of solute molecules can also be affected by the selectivity rate ($x$) of the selectivity filter in the pore.

$$\alpha = x(1-\sigma_s)$$

The pH sensitivity of AQP permeability was first investigated by Zeuthen & Klaerke[43], who found that as the pH value dropped from 7.4 to 6.4, AQP3 lost half of its water permeability reversibly. Besides AQP3, the water permeation of AQP0 [48] and AQP4[49] reach a maximum at pH 6.5 and pH 8.5 respectively, while AQP6[50] is activated below pH 5.5. From Equation (3), we can find that the driving force of AQP osmosis changes with the equivalent diameter of its vestibule. The equivalent diameter ($d$) of the vestibule surrounded by loops with abundant polar amino acids can be effected by the pH of the solution[48,51]. Then $\sigma_s$ of solutes to AQP will be changed. Therefore, only small changes in side-chain positions of loops are enough to regulate AQPs' water permeability rather than requiring a global change in its structure. This agrees with the x-ray structure of AQP0 by Harries and coworkers[52]: there is little evidence in the structure to support blockade in a static sense at pH 10. Therefore, an AQP may be able to work as a regulation valve by adjusting its vestibule size, not just as a block valve which only can open or close[53]. This possible rapid permeability regulating mechanism of AQPs may offer cells more flexibility to adapt various environments.

Separating efficiency and operating pressure are two important indicators of artificial membrane properties. The former has determinative effects on the purity of the purpose product while the second one has a direct relation to the energy consumption of the membrane separating system[54]. Unfortunately, from K-K equations, it is found that the performance improvements of these two indicators seem to be mutually incompatible goals. However, from the modified model in this work, we can find that high separating efficiency and low operating pressure can be achieved simultaneously if the pore has special geometry as AQP. The dumbbell-shaped structure of the AQPs makes them have both excellent water permeability and outstanding selectivity to solute molecules. This can inspire the design of new membranes with better performances.

## Acknowledgements

The authors thank A.E Hill, Peter Agre, and John Mathai for useful discussions and valuable suggestions through e-mail. Mr. and Mrs. Serody (Robert and Lucy), Miss Yibo Zhang are also gratefully acknowledged for reading our manuscript and giving useful comment. This work was supported by the National Natural Science Foundation of China (No. 51076057)

## Reference

[1] A.D. Khawaji, I.K. Kutubkhanah, J.M. Wie, Advances in seawater desalination technologies, Desalination. 221 (2008) 47–69. doi:10.1016/j.desal.2007.01.067.

[2] M. Elimelech, W.A. Phillip, The Future of Seawater Desalination: Energy, Technology, and the Environment, Science (80-. ). 333 (2011) 712–717. doi:10.1126/science.1200488.

[3] K.P. Lee, T.C. Arnot, D. Mattia, A review of reverse osmosis membrane materials for desalination—Development to date and future potential, J. Memb. Sci. 370 (2011) 1–22. doi:10.1016/j.memsci.2010.12.036.

[4] C. Christy, S. Vermant, The state-of-the-art of filtration in recovery processes for biopharmaceutical production, Desalination. 147 (2002) 1–4. doi:10.1016/S0011-9164(02)00562-3.

[5] R. van Reis, A. Zydney, Bioprocess membrane technology, J. Memb. Sci. 297 (2007) 16–50. doi:10.1016/j.memsci.2007.02.045.

[6] C. Smith, Striving for purity: advances in protein purification, Nat. Methods. 2 (2005) 71–77. doi:10.1038/nmeth0105-71.

[7] C. Charcosset, Membrane processes in biotechnology: An overview, Biotechnol. Adv. 24 (2006) 482–492. doi:10.1016/j.biotechadv.2006.03.002.

[8] S. Loeb, Energy production at the Dead Sea by pressure-retarded osmosis: challenge or chimera?, Desalination. 120 (1998) 247–262. doi:10.1016/S0011-9164(98)00222-7.

[9] S. Loeb, One hundred and thirty benign and renewable megawatts from Great Salt Lake? The possibilities of hydroelectric power by pressure-retarded osmosis, Desalination. 141 (2001) 85–91. doi:10.1016/S0011-9164(01)00392-7.

[10] S. Loeb, Large-scale power production by pressure-retarded osmosis, using river water and sea water passing through spiral modules, Desalination. 143 (2002) 115–122. doi:10.1016/S0011-9164(02)00233-3.

[11] A. Achilli, T.Y. Cath, A.E. Childress, Power generation with pressure retarded osmosis: An experimental and theoretical investigation, J. Memb. Sci. 343 (2009) 42–52. doi:10.1016/j.memsci.2009.07.006.

[12] A. Achilli, A.E. Childress, Pressure retarded osmosis: From the vision of Sidney Loeb to the first prototype installation - Review, Desalination. 261 (2010) 205–211. doi:10.1016/j.desal.2010.06.017.

[13] B.E. Logan, M. Elimelech, Membrane-based processes for sustainable power generation using water, Nature. 488 (2012) 313–319. doi:10.1038/nature11477.

[14] O. Kedem, A. Katchalsky, Thermodynamic analysis of the permeability of biological membranes to non-electrolytes, Biochim. Biophys. Acta. 27 (1958) 229–246. doi:10.1016/0006-3002(58)90330-5.

[15] K.S. Spiegler, O. Kedem, Thermodynamics of hyperfiltration (reverse osmosis): criteria for efficient membranes, Desalination. 1 (1966) 311–326. doi:10.1016/S0011-9164(00)80018-1.

[16] J.L. Anderson, D.M. Malone, Mechanism of Osmotic Flow in Porous Membranes, Biophys. J. 14 (1974) 957–982. doi:10.1016/S0006-3495(74)85962-X.

[17] J.L. Anderson, Configurational effect on the reflection coefficient for rigid solutes in capillary pores, J. Theor. Biol. 90 (1981) 405–426. doi:10.1016/0022-5193(81)90321-0.

[18] G. Bhalla, W.M. Deen, Effects of molecular shape on osmotic reflection coefficients, J. Memb. Sci. 306 (2007) 116–124. doi:10.1016/j.memsci.2007.08.025.

[19] S. Gravelle, L. Joly, F. Detcheverry, C. Ybert, C. Cottin-Bizonne, L. Bocquet, Optimizing water permeability through the hourglass shape of aquaporins., Proc. Natl. Acad. Sci. U. S. A. 110 (2013) 16367–72. doi:10.1073/pnas.1306447110.

[20] S. Gravelle, L. Joly, C. Ybert, L. Bocquet, Large permeabilities of hourglass nanopores: From hydrodynamics to single file transport, J. Chem. Phys. 141 (2014) 1–7. doi:10.1063/1.4897253.

[21] X. Gong, J. Li, H. Zhang, R. Wan, H. Lu, S. Wang, et al., Enhancement of water permeation across a nanochannel by the structure outside the channel, Phys. Rev. Lett. 101 (2008) 1–4. doi:10.1103/PhysRevLett.101.257801.

[22] J. Li, X. Gong, H. Lu, D. Li, H. Fang, R. Zhou, Electrostatic gating of a nanometer water channel., Proc. Natl. Acad. Sci. U. S. A. 104 (2007) 3687–3692. doi:10.1073/pnas.0604541104.

[23] T.B. Sisan, S. Lichter, The end of nanochannels, Microfluid. Nanofluidics. 11 (2011) 787–791. doi:10.1007/s10404-011-0855-9.

[24] H. Sui, B.G. Han, J.K. Lee, P. Walian, B.K. Jap, Structural basis of water-specific transport through the AQP1 water channel., Nature. 414 (2001) 872–878. doi:10.1038/414872a.

[25] T. Gonen, T. Walz, The structure of aquaporins., Q. Rev. Biophys. 39 (2006) 361–396. doi:10.1017/S0033583506004458.

[26] G.T. Rich, R.I. Sha'afi, T.C. Barton, a K. Solomon, Permeability studies on red cell membranes of dog, cat, and beef., J. Gen. Physiol. 50 (1967) 2391–2405.

[27] M.R. Toon, A.K. Solomon, Permeability and Reflection Coefficients of Urea and Small Amides in the Human Red Cell, J. Membr. Biol. 153 (1996) 137–146. doi:10.1007/s002329900117.

[28] I.S. Davis, B. Shachar-Hill, M.R. Curry, K.S. Kim, T.J. Pedley, a. E. Hill, Osmosis in semi-permeable pores: an examination of the basic flow equations based on an experimental and

molecular dynamics study, (2007). doi:10.1098/rspa.2006.1803.

[29] M.R. Curry, B. Shachar-Hill, A.E. Hill, Single Water Channels of Aquaporin-1 Do not Obey the Kedem-Katchalsky Equations, J. Membr. Biol. 181 (2001) 115–123. doi:10.1007/s00232-001-0015-3.

[30] T. Zeuthen, M. Alsterfjord, E. Beitz, N. Macaulay, Osmotic water transport in aquaporins: evidence for a stochastic mechanism., J. Physiol. 591 (2013) 5017–29. doi:10.1113/jphysiol.2013.261321.

[31] D. a Goldstein, a K. Solomon, Determination of equivalent pore radius for human red cells by osmotic pressure measurement, J. Gen. Physiol. 44 (1960) 1–17. http://www.pubmedcentral.nih.gov/articlerender.fcgi?artid=2195086&tool=pmcentrez&rendertype=abstract.

[32] J.H. van't Hoff, The role of osmotic pressure in the analogy between solutions and gases, J. Memb. Sci. 100 (1995) 39–44. doi:10.1016/0376-7388(94)00232-N.

[33] F. Kiil, Kinetic model of osmosis through semipermeable and solute-permeable membranes, Acta Physiol. Scand. 177 (2003) 107–117. doi:10.1046/j.1365-201X.2003.01062.x.

[34] W.G. Mcmillan, J.E. Mayer, The Statistical Thermodynamics of Multicomponent Systems, J. Chem. Phys. 13 (1945) 276–305. doi:10.1063/1.1724036.

[35] M. Knudsen, The Kinetic Theory of Gases. Some Modern Aspects, 1950.

[36] J.C. Mathai, S. Mori, B.L. Smith, G.M. Preston, N. Mohandas, M. Collins, et al., Functional Analysis of Aquaporin-1 Deficient Red Cells: THE COLTON-NULL PHENOTYPE, J. Biol. Chem. 271 (1996) 1309–1313. doi:10.1074/jbc.271.3.1309.

[37] A.H. Soll, A new approach to molecular configuration applied to aqueous pore transport., J. Gen. Physiol. 50 (1967) 2565–78. http://www.ncbi.nlm.nih.gov/pubmed/5584621.

[38] S. Azouzi, M. Gueroult, P. Ripoche, S. Genetet, Y.C. Aronovicz, C. Le Van Kim, et al., Energetic and molecular water permeation mechanisms of the human red blood cell urea transporter B, PLoS One. 8 (2013) 2–12. doi:10.1371/journal.pone.0082338.

[39] B. Yang, a. S. Verkman, Analysis of double knockout mice lacking aquaporin-1 and urea transporter UT-B. Evidence for UT-B-facilitated water transport in erythrocytes, J. Biol. Chem. 277 (2002) 36782–36786. doi:10.1074/jbc.M206948200.

[40] L. Shu, Y. Li, Xiaokang, L.X. Qian, S. Huang, S. Jin, et al., Aquaporin-1 is a Maxwell's Demon in the Body, (2015) 1–25. http://arxiv.org/abs/1511.07177.

[41] J. Masciocchi, G. Frau, M. Fanton, M. Sturlese, M. Floris, L. Pireddu, et al., MMsINC: a large-scale chemoinformatics database., Nucleic Acids Res. 37 (2009) D284–90. doi:10.1093/nar/gkn727.

[42] G.T. Rich, Permeability Studies on Red Cell Membranes of Dog, Cat, and Beef, J. Gen. Physiol. 50 (1967) 2391–2405. doi:10.1085/jgp.50.10.2391.

[43] T. Zeuthen, D.A. Klaerke, Transport of water and glycerol in aquaporin 3 is gated by H+, J. Biol. Chem. 274 (1999) 21631–21636. doi:10.1074/jbc.274.31.21631.

[44] T. Gonen, Y.F. Cheng, P. Sliz, Y. Hiroaki, Y. Fujiyoshi, S.C. Harrison, et al., Lipid-protein interactions in double-layered two-dimensional AQPO crystals, Nature. 438 (2005) 633–638. doi:10.1038/nature04321.

[45] G. Soveral, A. Madeira, M.C. Loureiro-Dias, T.F. Moura, Membrane tension regulates water transport in yeast, Biochim. Biophys. Acta - Biomembr. 1778 (2008) 2573–2579. doi:10.1016/j.bbamem.2008.07.018.

[46] M. Ozu, R. a. Dorr, F. Gutiérrez, M. Teresa Politi, R. Toriano, Human AQP1 is a constitutively open channel that closes by a membrane-tension-mediated mechanism, Biophys. J. 104 (2013) 85–95. doi:10.1016/j.bpj.2012.11.3818.

[47] L. Leitão, C. Prista, M.C. Loureiro-Dias, T.F. Moura, G. Soveral, The grapevine tonoplast aquaporin TIP2;1 is a pressure gated water channel, Biochem. Biophys. Res. Commun. 450 (2014) 289–294. doi:10.1016/j.bbrc.2014.05.121.

[48] K.L. Nemeth-Cahalan, J.E. Hall, pH and Calcium Regulate the Water Permeability of Aquaporin 0, J. Biol. Chem. 275 (2000) 6777–6782. doi:10.1074/jbc.275.10.6777.

[49] K.L. Nemeth-Cahalan, Molecular Basis of pH and Ca2+ Regulation of Aquaporin Water Permeability, J. Gen. Physiol. 123 (2004) 573–580. doi:10.1085/jgp.200308990.

[50] M. Yasui, A. Hazama, T.H. Kwon, S. Nielsen, W.B. Guggino, P. Agre, Rapid gating and anion permeability of an intracellular aquaporin., Nature. 402 (1999) 184–7. doi:10.1038/46045.

[51] C. Tournaire-Roux, M. Sutka, H. Javot, E. Gout, P. Gerbeau, D.-T. Luu, et al., Cytosolic pH regulates root water transport during anoxic stress through gating of aquaporins., Nature. 425 (2003) 393–397. doi:10.1038/nature01853.

[52] W.E.C. Harries, D. Akhavan, L.J.W. Miercke, S. Khademi, R.M. Stroud, The channel architecture of aquaporin 0 at a 2.2-A resolution., Proc. Natl. Acad. Sci. U. S. A. 101 (2004) 14045–50. doi:10.1073/pnas.0405274101.

[53] S. Törnroth-Horsefield, Y. Wang, K. Hedfalk, U. Johanson, M. Karlsson, E. Tajkhorshid, et al., Structural mechanism of plant aquaporin gating., Nature. 439 (2006) 688–694. doi:10.1038/nature04316.

[54] M.H.I. Dore, Forecasting the economic costs of desalination technology, Desalination. 172 (2005) 207–214. doi:10.1016/j.desal.2004.07.036.